\documentclass[aps,prl,twocolumn,superscriptaddress]{revtex4}
\usepackage{mathrsfs}
\usepackage{epsfig}
\usepackage{graphicx}
\usepackage{amsfonts}
\usepackage[figuresright]{rotating}
\usepackage{amssymb}
\usepackage{amsmath}
\usepackage{dcolumn}
\usepackage{bm}
\usepackage{color}

\def\be{\begin{equation}} \def\ee{\end{equation}}
\def\bea{\begin{eqnarray}} \def\eea{\end{eqnarray}}

\newcommand{\WQCASQC} {Wilczek Quantum Center and Key Laboratory of Artificial Structures and Quantum Control, Shanghai Research Center for Quantum Sciences, School of Physics and Astronomy, Shanghai Jiao Tong University, Shanghai 200240, China}

\begin{document}
\title{Quantum slush state in Rydberg atom arrays}

\author{Tengzhou Zhang}
\affiliation{\WQCASQC}

\author{Zi Cai}
\email{zcai@sjtu.edu.cn}
\affiliation{\WQCASQC}

\begin{abstract}  
In this study, we propose an exotic quantum state which does not order at zero temperature  in a Rydberg atom array with antiblockade mechanism. By performing an unbiased large-scale quantum Monte Carlo simulation, we investigate a minimal model with facilitated excitation in a disorder-free system. At zero temperature, this model exhibits a heterogeneous structure of liquid and glass mixture.  This state, dubbed quantum slush state,  features a quasi-long-range order with an algebraic decay for its correlation function, and is different from most well-established quantum phases of matter.

\end{abstract}


\maketitle

{\it Introduction --} The search for highly correlated quantum matters that do not order even at zero temperature has been a focus across several disciplines of quantum physics in the past decades. One of the best-studied examples in this regards are the quantum spin liquids which feature long-range quantum entanglement and topological excitations\cite{Wen2017,Zhou2017}. Recently, Rydberg atom arrays has provided promising neutral-atom platforms for quantum simulation\cite{Weimer2010,Semeghini2021,Ebadi2021,Ebadi2022}. A key element therein is the Rydberg blockade mechanism: double excitations of adjacent atoms are strongly suppressed\cite{Jaksch2000,Lukin2001}. The kinetic constraint, incorporating quantum fluctuations, can give rise to intriguing quantum matters in\cite{Samajdar2020,Samajdar2021,Verresen2021,Yue2021,Yan2023,Sfairopoulos2023,Sfairopoulos2024} and out\cite{Lesanovsky2013,Lan2018,Bluvstein2021,Wu2023} of equilibrium.

In this study, an exotic quantum liquid state other than the spin liquids is proposed, which is based on a facilitation (antiblockade) mechanism with a kinetic constraint\cite{Ates2007,Amthor2010}: an atom can be coherently excited if and only if just one of its neighbors has already been in the Rydberg state. In another word, an excited atom facilitates the excitation its neighbors. Unlike most work in this regards which mainly focus on the real time evolution of this model\cite{Garttner2013,Lesanovsky2014,Marcuzzi2017,Espigares2017,Causer2020,Helmrich2020,Ding2020,Liu2022}, here we study the equilibrium properties of a minimal model supporting such a kinetic constriant  using an unbiased quantum Monte Carlo (QMC) simulation\cite{Sandvik2003,Merali2021,Patil2023}.

Here, we show that even for a simple  disorder-free model without frustration, the interplay between the kinetic constraint and quantum fluctuation can give rise to an intriguing quantum state exhibits a heterogeneous structure of liquid and glass mixture. Similar dynamical heterogeneity was observed before in a classical frustrated system\cite{Rau2016}, and the corresponding classical spin liquid state is dubbed ``spin slush'' in analogy to the mixture of solid and liquid water.  The quantum slush state proposed in this study does not order at zero temperature, and features an algebraic decaying in its correlation function, distinguishing it from most well-established quantum phases of matter as well as the classical spin slush state.

\begin{figure}[htb]
\includegraphics[width=0.99\linewidth]{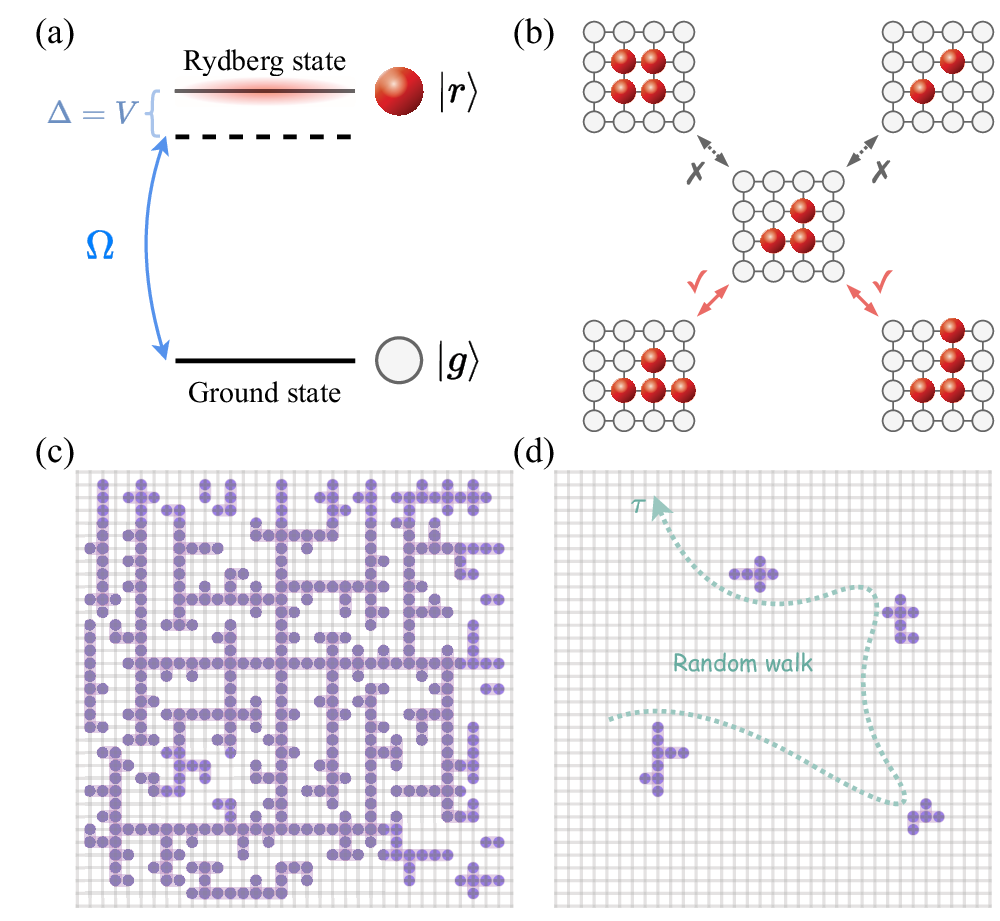}
\caption{(Color online) (a)Each atom has two internal states ($|g\rangle$ and $|r\rangle$),  $|g\rangle$ can be coherently excited to $|r\rangle$ by a laser with Rabi frequency $\Omega$ and detuning $\Delta$. The transition  becomes resonant if the detuning compensates the NN interaction ($\Delta=V$); (b)Examples of allowable (the lower two panels) and forbidden (the upper two panels) transitions between different configurations under the kinetic constraint. (c)A typical configuration generated by the importance sampling in QMC simulation for the quantum slush phase with $\mu=0$. (d)A trajectory of a finite cluster generated by QMC simulation in the mobile finite cluster phase with $\mu=-1.1J$. } \label{fig:fig1}
\end{figure}

{\it Model --} The system we studied is a Rydberg array where the atoms are placed in a two-dimensional (2D) square lattice. Each atom possesses two internal states: a ground state $|g\rangle$ can be coherently excited to a Rydberg state $|r\rangle$ by a laser with Rabi frequency $\Omega$ and detuning $\Delta$. The interaction strength between a pair of nearest-neighboring (NN) excited atoms is $V$, and we ignore the longer-range interaction for simplicity. The facilitated excitation emerges if the detuning happens to be compensated by the NN interaction ($\Delta=V$), where the coherent transition between $|g\rangle$ and $|r\rangle$ at a particular site becomes resonant only if just one of its neighbors is already in state $|r\rangle$ (see Fig. \ref{fig:fig1}).  This anti-blockade  mechanism has been realized in Rydberg aggregates, where the strongly correlated growth induced by this facilitated excitation has been explored experimentally\cite{Schempp2014,Urvoy2015}.

Throughout this study, we ignore the incoherent transition and dissipation, and the system can be considered as a closed system.   The minimal model describing the resulting quantum many-body system is rather simple, and its Hamiltonian reads:
\begin{equation}
H=\sum_i [-J \Xi_\mathbf{i} \sigma_\mathbf{i}^x-\mu n_\mathbf{i}], \label{eq:Ham}
\end{equation}
where $J$ and $\mu$ represent the amplitude of the coherent transition and the effective detuning respectively. $\hat{\sigma}^x_\mathbf{i}=|g\rangle_\mathbf{i}\langle r|+|r\rangle_\mathbf{i}\langle g|$ and $n_\mathbf{i}=|r\rangle_\mathbf{i}\langle r|$ is the  density operator of the Rydberg state on site $\mathbf{i}$. $\Xi_\mathbf{i}$ is the operator projecting onto the subspace with only one of the four neighbors of site $\mathbf{i}$ being in  $|r\rangle$ state. $\Xi_\mathbf{i}$ can be expressed in terms of the density operators of its neighbors as:
\begin{equation}
\Xi_i=\sum_{\alpha} \big{[}n_{\mathbf{i}+\mathbf{e}_\alpha}\prod_{\gamma \neq \alpha}(1-n_{\mathbf{i}+\mathbf{e}_\gamma})\big{]} \label{eq:constraint}
\end{equation}
where $\alpha,\gamma=1\sim 4$, $\mathbf{e}_{1\sim 4}$ indicate the unit vectors along $\pm x$ and $\pm y$ directions and $\mathbf{i}+\mathbf{e}_{1\sim 4}$ represent the four neighbors of site $\mathbf{i}$. It is easy to check that $\Xi_i=1$ if and only if $\sum_{\alpha}n_{\mathbf{i}+\mathbf{e}_\alpha}=1$, otherwise $\Xi_i=0$. An one-dimensional (1D) analogue of this model has been investigated numerically\cite{Causer2020,Zadnik2023}.

The kinetic constraint imposes additional conserved quantities on the system. For each configuration, we define  {\it a cluster} as a group of connected sites occupied by $|r\rangle$ atoms, and the total number of these clusters is conserved since the spin flip terms under the kinetic constraint does not change the connectivity of a cluster ({\it e.g.} they cannot break one cluster into two, see Fig.\ref{fig:fig1} b).  This conserved quantity further divides the total Hilbert space into different subspaces, each of which is characterized by its cluster number.  In the following, we focus on the simplest subspace with only one cluster, and general situations will be discussed in the Supplementary material\cite{Supplementary}. We will argue that even though the lowest energy state within the one-cluster sector is not the true groundstate of Ham.(\ref{eq:Ham}), it can be considered as a metastable state that can be observed in realistic Rydberg experimental setup.

{\it Method --} We implement the QMC simulation with the stochastic series expansion (SSE) algorithm\cite{Sandvik2003,Merali2021,Patil2023} on a system in a $L\times L$ square lattice. The periodical boundary condition (PBC) is chosen and the inverse temperature is set as $\beta=L$ to scale to the lowest energy state in the thermodynamic limit. Notably,  the SSE algorithm also preserves the cluster number. In our QMC simulation, we focus on the single-cluster subspace by choosing the initial state with only one ``seed'': a configuration with only one site is in $|r\rangle$ and all others are in $|g\rangle$. This cluster will keep growing and ramifying in the QMC simulation. It is straightforward to implement the kinetic constraint in the SSE algorithm. Since this constraint as shown in Eq.(\ref{eq:constraint}) occurs in Fock basis of the particle number representation, it does not cause negative sign problem during the QMC sampling. As a result, the QMC simulation can give reliable results for systems of sufficiently large size and low temperature.

In the QMC simulation, the space-time configurations are sampled according to their weight in the partition function. To calculate the average value of operators which is diagonal in the Fock basis, one can choose the spacial configurations at different (imaginary) time slices, calculate the corresponding expectation values, then perform the average over  them. Two typical configurations generated in the QMC simulation for different $\mu$ have been shown in Fig.\ref{fig:fig1} (c) and (d), which are qualitatively different from each other and suggest a phase transition between them, as we will analyzed latter. For large $\mu$, the size of the single cluster is compatible to the system size, and it exhibits loopless branching structure as shown in Fig.\ref{fig:fig1} c. Similar structures widely exist in nature, such as lungs, coral, polymers, etc and  also appear in dynamical processes such as particle aggregation, crystal growth and electrochemical decomposition,  which have been shown to produce fractal clusters\cite{Vicsek1992}. The emergent dynamical fractal has also been used to explain the anomalous noise observed in a clean spin ice system\cite{Hallen2022}. The fractal dimension of the typical configurations in our case can be measured via the cluster-growing method\cite{Supplementary}, which has been used to characterize the fractal dimension for the loopless branching structure\cite{Vicsek1992,Feder1988}.

\begin{figure}[htb]
\includegraphics[width=0.9\linewidth]{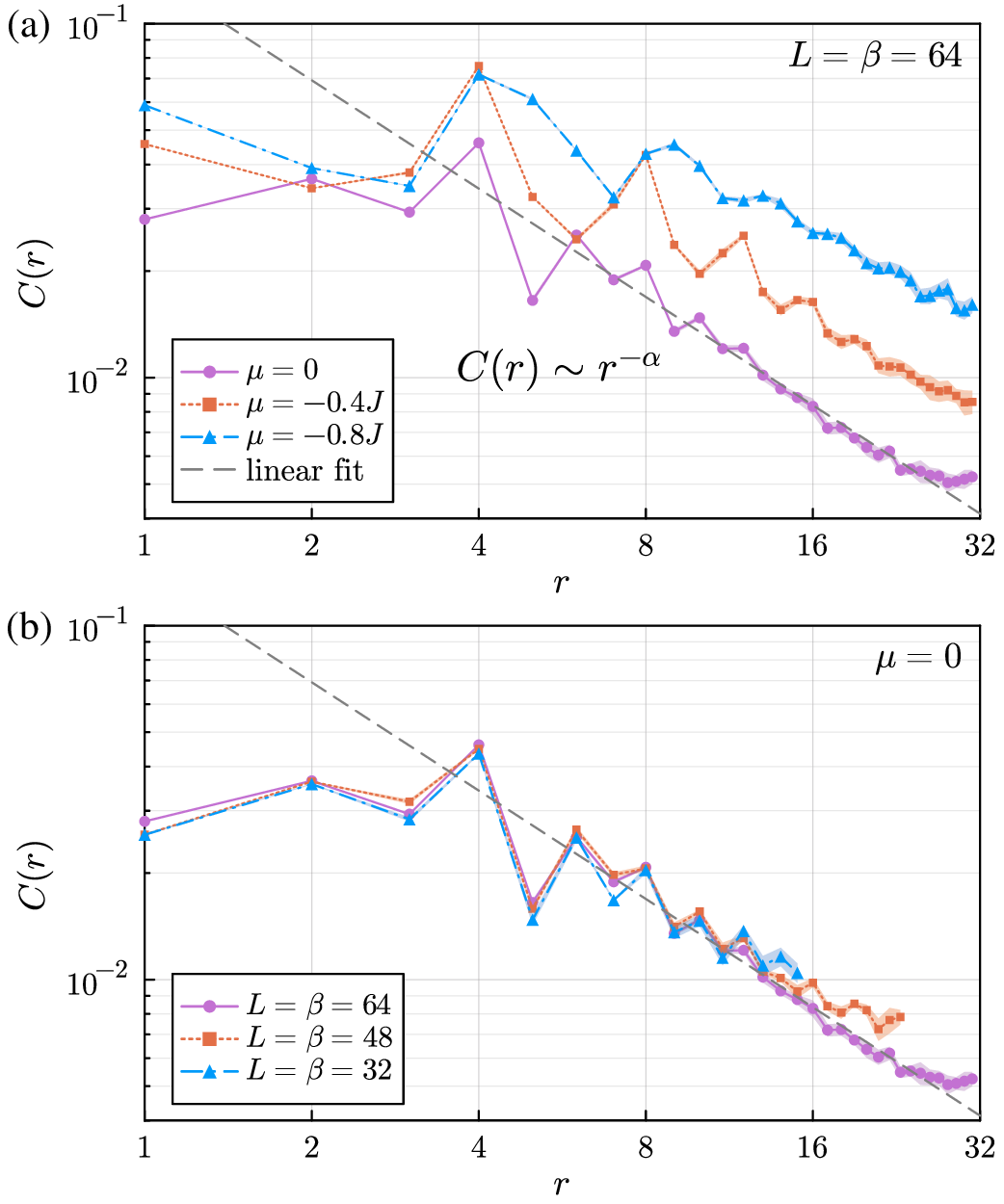}
\caption{(Color online) (a) The correlation functions $C(\mathbf{r})$ along the x-direction $\mathbf{r}=(r,0)$ for the system with different $\mu$ ($L=\beta=64$), each of which exhibits algebraic decay with the same exponent $\alpha$; (b) $C(\mathbf{r})$ in systems with a fixed $\mu=0$ but different $L$ and $\beta$.} \label{fig:fig2}
\end{figure}

\begin{figure}[htb]
\includegraphics[width=0.99\linewidth]{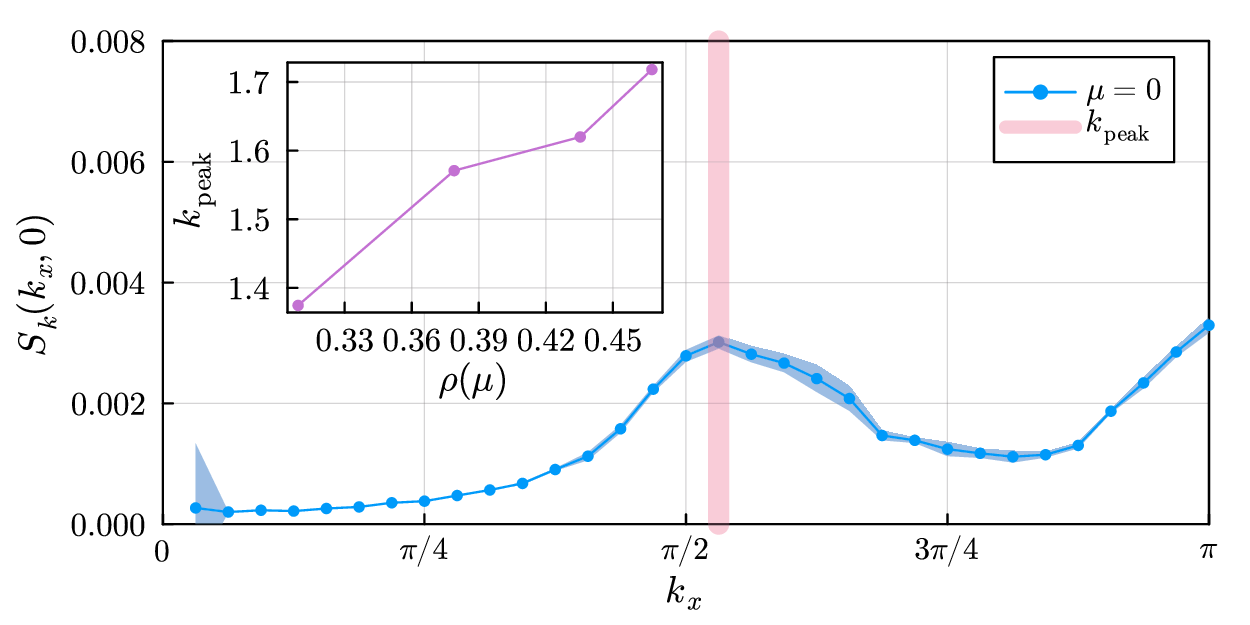}
\caption{(Color online) Structure factor along x-direction $S(k_x,0)$, which exhibits a peak at $k_x=k_{peak}$. The inset indicates the dependence of the peak position on the average density of the $|r\rangle$ state. ($L=\beta=64$ and $\mu=0$).} \label{fig:fig3}
\end{figure}

{\it Spatial correlation function: a quasi-long-range order with algebraic decay --} Here we focus on the properties of the lowest-energy state within the one-cluster sector for the case with $\mu>-J$, while leave the case with $\mu<-J$ for latter discussion.  The typical configurations in this case exhibit fractal feature, which is usually characterized by a power-law behavior.   To numerically validate this aspect, we calculate the correlation function on this state:
\begin{equation}
C(\mathbf{r})=\frac{1}{L^2}\sum_\mathbf{i}[\langle n_\mathbf{i} n_{\mathbf{i}+\mathbf{r}}\rangle-\langle n_\mathbf{i}\rangle\langle n_j\rangle]
\end{equation}
as well as the corresponding structure factor:
\begin{equation}
S(\mathbf{k})=\frac 1{L^2}\sum_\mathbf{r} C(\mathbf{r}) e^{i \mathbf{r}\cdot\mathbf{k}}
\end{equation}
In Fig.\ref{fig:fig2}, we plot $C(\mathbf{r})$ along x-direction, which exhibits an algebraic decay accompanied by an oscillation. One can fit the exponent of the algebraic decay as $\alpha=0.98$ for $\mu=0$, and it doesn't change significantly for different $\mu$.   However, due to the statistical error bar, the oscillating feature of the decaying as well as the finite size effect, it is difficult to conclude that whether this exponent $\alpha$ is independent on $\mu$, and whether it takes an integer value $\alpha=1$ even though it is close to it. Such a power-law correlation reminds us of the algebraic decay in the Coulomb phase of classical spin ice, which also takes integer values ($\alpha=d$ with d being the system dimension). However, our state is a pure quantum state thus is different from the classical Coulomb phase in the spin ice. A long-range order usually indicates a sharp peak in the structure factor, with its height diverging in the thermodynamic limit. However, Fig.~\ref{fig:fig3} reveals no such sharp peaks in $S(\mathbf{k})$. Instead, $S(\mathbf{k})$ exhibits a broad peak at the wavevector $k_{peak}$, which corresponds to the oscillation period in $C(r)$. We also observe that $k_{peak}$ grows with the average density $\rho=\frac 1{L^2} \sum_i \langle n_\mathbf{i}\rangle$ (see the inset of Fig.~\ref{fig:fig3}).

\begin{figure}[htb]
\includegraphics[width=0.99\linewidth]{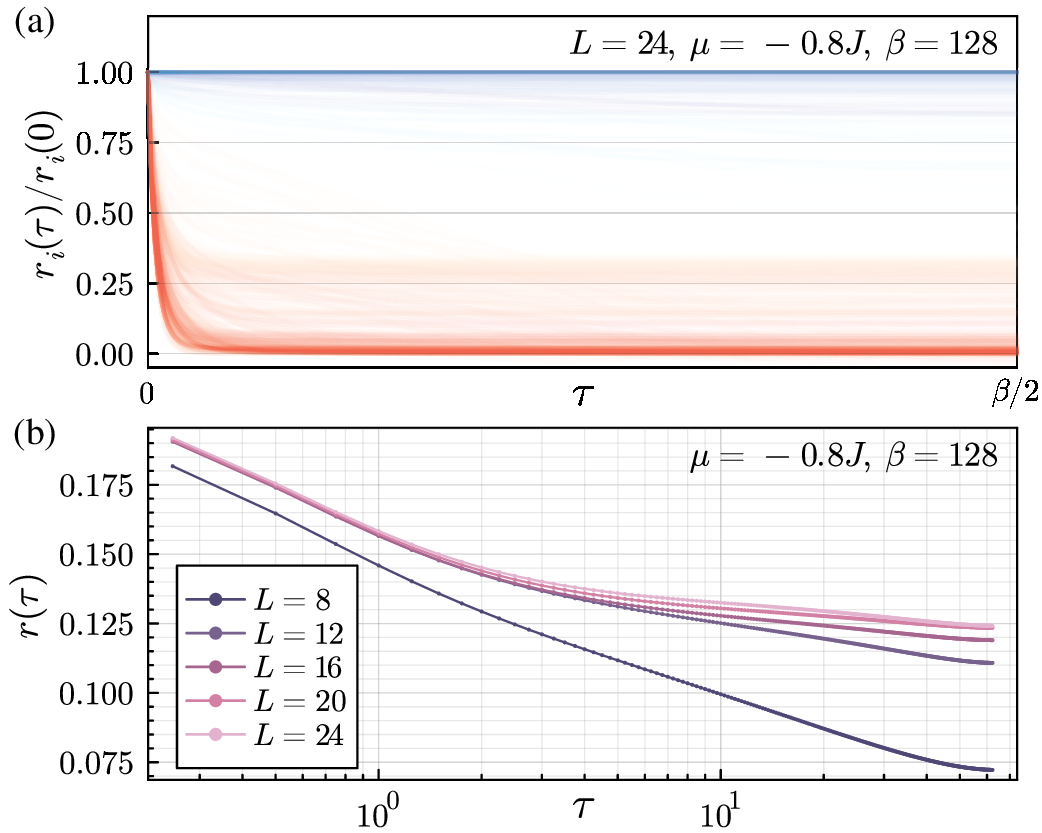}
\caption{(Color online) (a) The normalized autocorrelation functions $r_i(\tau)/r_i(0)$ for the systems with the parameters $\mu=-0.8J$, $L=24$ and $\beta=128J^{-1}$ (different curves indicate $r_i(\tau)$ on different sites.) (b)The averaged autocorrelation function for systems with different system size $L$. $\mu=-0.8J$ and $\beta=128J^{-1}$ } \label{fig:fig4}
\end{figure}

{\it Temporal correlation function:  a dynamical heterogeneity of liquid and glass mixture --} The typical configurations in our model exhibit branching network structure as shown in Fig.\ref{fig:fig1} (c), while the quantum fluctuations (the spin flipped terms) create a superposition among them. For a given configuration, the kinetic constraint dictates that the spin flipped terms can only operate on endpoints of the cluster, while the interior points of the cluster are immune to quantum fluctuations. This robustness against spin flipping indicates a glassy behavior with slow relaxation dynamics for those interior points of the network.  For such a site,  in spite of its robustness against single-spin flip, it can slowly relax by transforming an interior point into a endpoint via a sequence of spin flipping induced by quantum fluctuations.

To characterize the dynamical heterogeneity of these two types of sites (endpoints and interior points), we calculate the on-site autocorrelation function $r(\tau)$ in imaginary time $\tau$ on different system sites:
\begin{equation}
r_i(\tau)=\langle (n_i(0)-\rho)(n_i(\tau)-\rho)\rangle_{\tau}
\end{equation}
where the average $\langle O\rangle_\tau$ is performed over a single trajectory of QMC simulation.  For a given site, if the quantum fluctuation is completely frozen, $r_i(\tau\rightarrow\infty)/r_i(0)\rightarrow 1$. Conversely, for those sites whose spins are frequently flipped, $r_i(\tau)$ will decay to zero. The autocorrelation functions on different sites are illustrated in Fig.\ref{fig:fig4} (a), from which we can find that $r_i(\tau)/r_i(0)$ decays with $\tau$ for the majority of system sites, while for others, quantum fluctuations remain frozen up to the imaginary time scale ($\tau=\beta/2$) in our simulation, indicating a glassy behavior. However, after sufficiently long time, those ``frozen'' sites will be eventually flipped via the multispin flipping processes analyzed above, which can be reflected in the averaged autocorrelation function $r(\tau)=\frac 1{L^2}\sum_i r_i(\tau)$ as shown in Fig.\ref{fig:fig4}.  $r(\tau)$ keeps decaying with $\tau$ with a decay rate that decreases with $L$ (b). As shown in the SM\cite{Supplementary}, the relaxation time for those ``frozen'' sites will grows exponentially with the system size, which agrees with the glassy dynamics.

Compared to the spin glass, our model is free from disorder. The glassy mechanism in such a translational invariant system can be attributed to the kinetic constraints, akin to those observed in genuine glassy systems\cite{Garrahan2018}. However, the quantum fluctuation (spin flipped terms operating on the unfrozen sites) enhances the coherence thus is responsible for the distinctive quasi-long-range order (algebraic instead of exponential decay of $C(\mathbf{r})$), which makes it qualitatively differ from the conventional glass phases with only short-range order. Such a heterogeneous structure of liquid and glass mixture is reminiscent  of an intriguing classical spin liquid state  dubbed ``spin slush'', which was first proposed by Rau {\it et.al} in a classical spin ice system\cite{Rau2016}.

\begin{figure}[htb]
\includegraphics[width=0.9\linewidth]{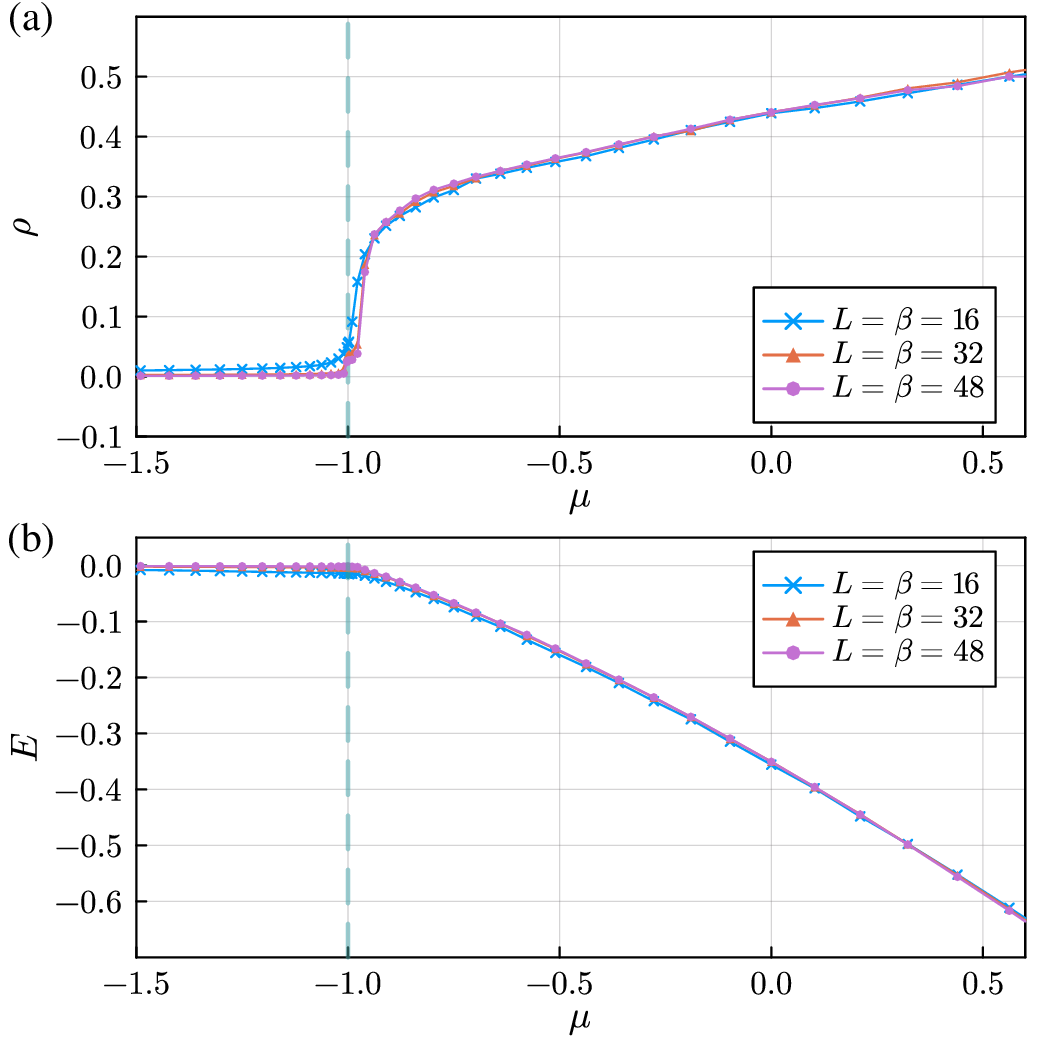}
\caption{(Color online) (a) The average density of the $|r\rangle$ states $\rho$ and (b) the ground state energy $E$ as a function of the detuning $\mu$ with different system size and $L=\beta$.} \label{fig:fig5}
\end{figure}

\begin{figure}[htb]
\includegraphics[width=0.99\linewidth]{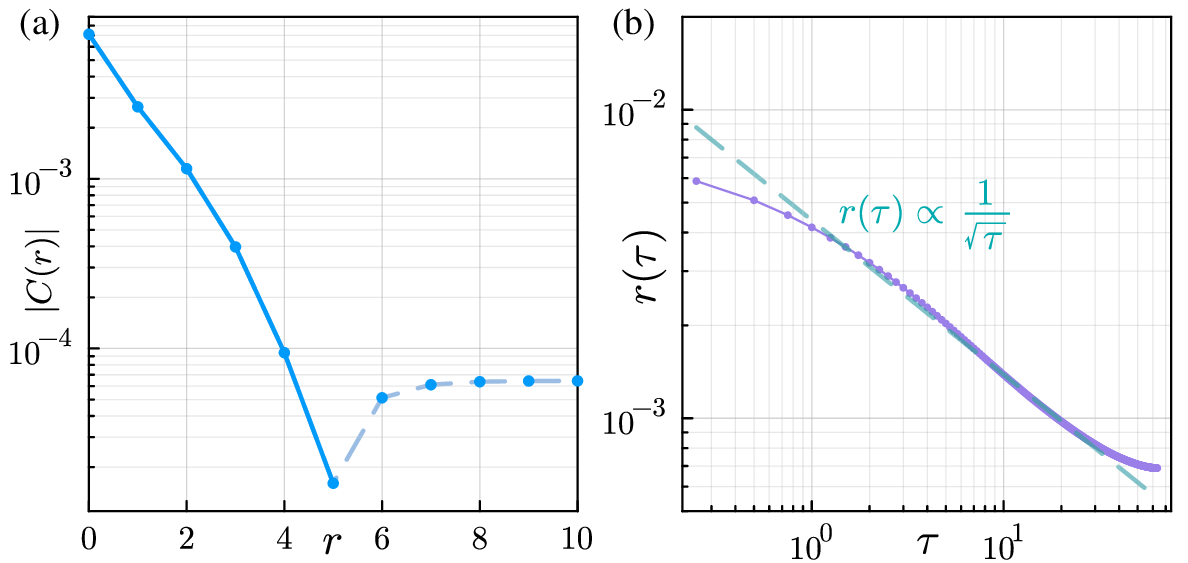}
\caption{(Color online) (a) The spatial correlation (along x-direction) and (b) the averaged autocorrelation function in the mobile finite-cluster phase with parameter $L=24$, $\beta=128J^{-1}$ and $\mu=-1.1J$. } \label{fig:fig6}
\end{figure}

{\it Quantum phase transition from a quantum slush to a mobile finite-size cluster state --} Fig. \ref{fig:fig1} (c) and (d) suggest there exists two different phases depending on the value of $\mu$. To characterize the phase transition between them, we calculate the average density $\rho$ as a function of $\mu$. As illustrated in Fig. \ref{fig:fig5} (a), there exists a critical value  at $\mu_c\approx -0.983J$, above which $\rho$ increases monotonically with $\mu$, and there is no plateau in the $\rho-\mu$ curve, indicating the absence of a Mott insulator. The ground state energy $E$ as a function of $\mu$ as shown in Fig. ~\ref{fig:fig5} (b) seems to imply a 1st order quantum phase transition  at $\mu=\mu_c$, which is akin to the dynamical large deviations transition observed in the 1D  models\cite{Causer2020,Zadnik2023}. However, due to the finite-site effect as well as the statistical error bar of the numerical results, it is difficult to preclude the possibility of a continuous quantum phase transition.

For a sufficiently low $\mu$, the energy cost of exciting an atom is so substantial that the cluster cannot grow. As a consequence, the average density $\rho\sim \frac{1}{L^2}$ thus will vanish  in the thermodynamic limit. This wavefunction of this state is a superposition of those Fock states containing only one finite-size cluster, whose size is determined by the value of $\mu$. This picture can be verified numerically from its spatial and temporal correlation function as shown in Fig.\ref{fig:fig6}. The spatial correlation function  decays exponentially in distance $C(r)\sim e^{-r/l_0}$, where the characteristic length scale $l_0$ reflects the average ``radius'' of the cluster and is independent on the system size. Different from the quantum slush state, here the averaged autocorrelation decays algebraically in time as  $r(\tau)\sim \tau^{-\frac 12}$, a  signature of the random walk for a single cluster in a 2D lattice.

{\it Discussion --} Throughout this paper, we focus on a specific subspace of Hilbert space with only one cluster. Although our model enforces a stringent kinetic constraint, real experimental systems implement this constraint through an energy penalty mechanism, which allows for potential violations in principle. Such a constraint-violating perturbation may induce quantum tunneling between subspaces characterized by different cluster numbers, and will transforms the previously studied state with a single cluster into a metastable state. The lifetime of this metastable state is inversely proportional to the strength of the perturbation\cite{Supplementary}. Consequently, as long as the lifetime of this proposed state greatly exceeds the typical time scale of a practical Rydberg atom system, it remains observable, permitting experimental investigation into its properties.

One may wonder whether the glassy dynamics on a few proportion of lattice sites  prevent the system from being thermalized, and make the system be trapped into an initial state-dependent local minimum state. To clarify this problem, we perform the ensemble average over a large number of QMC trajectories, and carefully compare  the average over different QMC trajectories and over the QMC simulation time. The results from these two average procedures agree with each other within the statistical error bar\cite{Supplementary}.  In addition, to check the initial state dependence of our result, we compare the results starting from two different initial states, and the difference between them is also sufficiently small\cite{Supplementary}. The equivalence between the ensemble and time averages as well as the initial state independence implies thermalization in our simulation.

{\it Conclusion and outlook --} In conclusion, the proposed  quantum slush state  is different from most well-established quantum phases of matter. It is neither an ordered phase, nor a Mott insulator. The presence of quasi-long-range order distinguishes it from conventional spin\cite{Mezard1986} and Bose glasses\cite{Giamarchi1988,Fisher1989}. This phase also stands apart from the U(1) quantum spin liquid\cite{Hermele2004} or the Coulomb phase in the classical spin ice\cite{Henley2010,Castelnovo2012} due to the absence of emergent gauge field.  It is not intrinsically tied to criticality, and exists in a wide regime  instead of at  a particular point in the parameter space.

One of the unsolved problems worthy of further study is the physical origin of the algebraic decay and the associated simple power exponent. In the Coulomb phase of spin ice, the algebraic decay can be understood as a consequence of the constraint imposed by the ice rule\cite{Henley2010}, one may wonder whether there exists a similar mechanism associated with kinetic constraint in our model.   Another unexplored avenue  is the effect of thermal fluctuation, for instance,  whether this quasi-long-range order can persist at low temperature analogous to the two-dimensional XY model\cite{Kosterlitz1973}, or is not robust against any thermal fluctuation like the Coulomb phase in the spin ice. A related question is what's the nature of the elementary excitations for such a state, do they manifest as collective excitations or an individual ones?

{\it Acknowledgments}.--- We acknowledge the helpful discussions with J. P. Garrahan and D. Huse.  This work is supported by the National Key Research and Development Program of China (Grant No. 2020YFA0309000), NSFC of  China (Grant No.12174251), Natural Science Foundation of Shanghai (Grant No.22ZR142830),  Shanghai Municipal Science and Technology Major Project (Grant No.2019SHZDZX01).


\begin{thebibliography}{48}
\expandafter\ifx\csname natexlab\endcsname\relax\def\natexlab#1{#1}\fi
\expandafter\ifx\csname bibnamefont\endcsname\relax
  \def\bibnamefont#1{#1}\fi
\expandafter\ifx\csname bibfnamefont\endcsname\relax
  \def\bibfnamefont#1{#1}\fi
\expandafter\ifx\csname citenamefont\endcsname\relax
  \def\citenamefont#1{#1}\fi
\expandafter\ifx\csname url\endcsname\relax
  \def\url#1{\texttt{#1}}\fi
\expandafter\ifx\csname urlprefix\endcsname\relax\def\urlprefix{URL }\fi
\providecommand{\bibinfo}[2]{#2}
\providecommand{\eprint}[2][]{\url{#2}}

\bibitem[{\citenamefont{Wen}(2017)}]{Wen2017}
\bibinfo{author}{\bibfnamefont{X.-G.} \bibnamefont{Wen}},
  \bibinfo{journal}{Rev. Mod. Phys.} \textbf{\bibinfo{volume}{89}},
  \bibinfo{pages}{041004} (\bibinfo{year}{2017}).

\bibitem[{\citenamefont{Zhou et~al.}(2017)\citenamefont{Zhou, Kanoda, and
  Ng}}]{Zhou2017}
\bibinfo{author}{\bibfnamefont{Y.}~\bibnamefont{Zhou}},
  \bibinfo{author}{\bibfnamefont{K.}~\bibnamefont{Kanoda}}, \bibnamefont{and}
  \bibinfo{author}{\bibfnamefont{T.-K.} \bibnamefont{Ng}},
  \bibinfo{journal}{Rev. Mod. Phys.} \textbf{\bibinfo{volume}{89}},
  \bibinfo{pages}{025003} (\bibinfo{year}{2017}).

\bibitem[{\citenamefont{Weimer et~al.}(2010)\citenamefont{Weimer, Muller,
  Lesanovsky, Zoller, and Buchler}}]{Weimer2010}
\bibinfo{author}{\bibfnamefont{H.}~\bibnamefont{Weimer}},
  \bibinfo{author}{\bibfnamefont{M.}~\bibnamefont{Muller}},
  \bibinfo{author}{\bibfnamefont{I.}~\bibnamefont{Lesanovsky}},
  \bibinfo{author}{\bibfnamefont{P.}~\bibnamefont{Zoller}}, \bibnamefont{and}
  \bibinfo{author}{\bibfnamefont{H.~P.} \bibnamefont{Buchler}},
  \bibinfo{journal}{Nature Phys.} \textbf{\bibinfo{volume}{6}},
  \bibinfo{pages}{382} (\bibinfo{year}{2010}).

\bibitem[{\citenamefont{Semeghini et~al.}(2021)\citenamefont{Semeghini, Levine,
  Keesling, Ebadi, Wang, Bluvstein, Verresen, Pichler, Kalinowski, Samajdar
  et~al.}}]{Semeghini2021}
\bibinfo{author}{\bibfnamefont{G.}~\bibnamefont{Semeghini}},
  \bibinfo{author}{\bibfnamefont{H.}~\bibnamefont{Levine}},
  \bibinfo{author}{\bibfnamefont{A.}~\bibnamefont{Keesling}},
  \bibinfo{author}{\bibfnamefont{S.}~\bibnamefont{Ebadi}},
  \bibinfo{author}{\bibfnamefont{T.~T.} \bibnamefont{Wang}},
  \bibinfo{author}{\bibfnamefont{D.}~\bibnamefont{Bluvstein}},
  \bibinfo{author}{\bibfnamefont{R.}~\bibnamefont{Verresen}},
  \bibinfo{author}{\bibfnamefont{H.}~\bibnamefont{Pichler}},
  \bibinfo{author}{\bibfnamefont{M.}~\bibnamefont{Kalinowski}},
  \bibinfo{author}{\bibfnamefont{R.}~\bibnamefont{Samajdar}},
  \bibnamefont{et~al.}, \bibinfo{journal}{Science}
  \textbf{\bibinfo{volume}{374}}, \bibinfo{pages}{1242} (\bibinfo{year}{2021}).

\bibitem[{\citenamefont{Ebadi et~al.}(2021)\citenamefont{Ebadi, Wang, Levine,
  Keesling, Semeghini, Omran, Bluvstein, Samajdar, Pichler, Ho
  et~al.}}]{Ebadi2021}
\bibinfo{author}{\bibfnamefont{S.}~\bibnamefont{Ebadi}},
  \bibinfo{author}{\bibfnamefont{T.~T.} \bibnamefont{Wang}},
  \bibinfo{author}{\bibfnamefont{H.}~\bibnamefont{Levine}},
  \bibinfo{author}{\bibfnamefont{A.}~\bibnamefont{Keesling}},
  \bibinfo{author}{\bibfnamefont{G.}~\bibnamefont{Semeghini}},
  \bibinfo{author}{\bibfnamefont{A.}~\bibnamefont{Omran}},
  \bibinfo{author}{\bibfnamefont{D.}~\bibnamefont{Bluvstein}},
  \bibinfo{author}{\bibfnamefont{R.}~\bibnamefont{Samajdar}},
  \bibinfo{author}{\bibfnamefont{H.}~\bibnamefont{Pichler}},
  \bibinfo{author}{\bibfnamefont{W.~W.} \bibnamefont{Ho}},
  \bibnamefont{et~al.}, \bibinfo{journal}{Nature}
  \textbf{\bibinfo{volume}{595}}, \bibinfo{pages}{227} (\bibinfo{year}{2021}).

\bibitem[{\citenamefont{Ebadi et~al.}(2022)\citenamefont{Ebadi, Keesling, Cain,
  Wang, Levine, Bluvstein, Semeghini, Omran, Liu, Samajdar et~al.}}]{Ebadi2022}
\bibinfo{author}{\bibfnamefont{S.}~\bibnamefont{Ebadi}},
  \bibinfo{author}{\bibfnamefont{A.}~\bibnamefont{Keesling}},
  \bibinfo{author}{\bibfnamefont{M.}~\bibnamefont{Cain}},
  \bibinfo{author}{\bibfnamefont{T.~T.} \bibnamefont{Wang}},
  \bibinfo{author}{\bibfnamefont{H.}~\bibnamefont{Levine}},
  \bibinfo{author}{\bibfnamefont{D.}~\bibnamefont{Bluvstein}},
  \bibinfo{author}{\bibfnamefont{G.}~\bibnamefont{Semeghini}},
  \bibinfo{author}{\bibfnamefont{A.}~\bibnamefont{Omran}},
  \bibinfo{author}{\bibfnamefont{J.}~\bibnamefont{Liu}},
  \bibinfo{author}{\bibfnamefont{R.}~\bibnamefont{Samajdar}},
  \bibnamefont{et~al.}, \bibinfo{journal}{Science}
  \textbf{\bibinfo{volume}{376}}, \bibinfo{pages}{1209} (\bibinfo{year}{2022}).

\bibitem[{\citenamefont{Jaksch et~al.}(2000)\citenamefont{Jaksch, Cirac,
  Zoller, Rolston, C\^ot\'e, and Lukin}}]{Jaksch2000}
\bibinfo{author}{\bibfnamefont{D.}~\bibnamefont{Jaksch}},
  \bibinfo{author}{\bibfnamefont{J.~I.} \bibnamefont{Cirac}},
  \bibinfo{author}{\bibfnamefont{P.}~\bibnamefont{Zoller}},
  \bibinfo{author}{\bibfnamefont{S.~L.} \bibnamefont{Rolston}},
  \bibinfo{author}{\bibfnamefont{R.}~\bibnamefont{C\^ot\'e}}, \bibnamefont{and}
  \bibinfo{author}{\bibfnamefont{M.~D.} \bibnamefont{Lukin}},
  \bibinfo{journal}{Phys. Rev. Lett.} \textbf{\bibinfo{volume}{85}},
  \bibinfo{pages}{2208} (\bibinfo{year}{2000}).

\bibitem[{\citenamefont{Lukin et~al.}(2001)\citenamefont{Lukin, Fleischhauer,
  Cote, Duan, Jaksch, Cirac, and Zoller}}]{Lukin2001}
\bibinfo{author}{\bibfnamefont{M.~D.} \bibnamefont{Lukin}},
  \bibinfo{author}{\bibfnamefont{M.}~\bibnamefont{Fleischhauer}},
  \bibinfo{author}{\bibfnamefont{R.}~\bibnamefont{Cote}},
  \bibinfo{author}{\bibfnamefont{L.~M.} \bibnamefont{Duan}},
  \bibinfo{author}{\bibfnamefont{D.}~\bibnamefont{Jaksch}},
  \bibinfo{author}{\bibfnamefont{J.~I.} \bibnamefont{Cirac}}, \bibnamefont{and}
  \bibinfo{author}{\bibfnamefont{P.}~\bibnamefont{Zoller}},
  \bibinfo{journal}{Phys. Rev. Lett.} \textbf{\bibinfo{volume}{87}},
  \bibinfo{pages}{037901} (\bibinfo{year}{2001}).

\bibitem[{\citenamefont{Samajdar et~al.}(2020)\citenamefont{Samajdar, Ho,
  Pichler, Lukin, and Sachdev}}]{Samajdar2020}
\bibinfo{author}{\bibfnamefont{R.}~\bibnamefont{Samajdar}},
  \bibinfo{author}{\bibfnamefont{W.~W.} \bibnamefont{Ho}},
  \bibinfo{author}{\bibfnamefont{H.}~\bibnamefont{Pichler}},
  \bibinfo{author}{\bibfnamefont{M.~D.} \bibnamefont{Lukin}}, \bibnamefont{and}
  \bibinfo{author}{\bibfnamefont{S.}~\bibnamefont{Sachdev}},
  \bibinfo{journal}{Phys. Rev. Lett.} \textbf{\bibinfo{volume}{124}},
  \bibinfo{pages}{103601} (\bibinfo{year}{2020}).

\bibitem[{\citenamefont{Samajdar et~al.}(2021)\citenamefont{Samajdar, Ho,
  Pichler, Lukin, and Sachdev}}]{Samajdar2021}
\bibinfo{author}{\bibfnamefont{R.}~\bibnamefont{Samajdar}},
  \bibinfo{author}{\bibfnamefont{W.~W.} \bibnamefont{Ho}},
  \bibinfo{author}{\bibfnamefont{H.}~\bibnamefont{Pichler}},
  \bibinfo{author}{\bibfnamefont{M.~D.} \bibnamefont{Lukin}}, \bibnamefont{and}
  \bibinfo{author}{\bibfnamefont{S.}~\bibnamefont{Sachdev}},
  \bibinfo{journal}{Proc. Nat. Acad. Sci.} \textbf{\bibinfo{volume}{118}},
  \bibinfo{pages}{e2015785118} (\bibinfo{year}{2021}).

\bibitem[{\citenamefont{Verresen et~al.}(2021)\citenamefont{Verresen, Lukin,
  and Vishwanath}}]{Verresen2021}
\bibinfo{author}{\bibfnamefont{R.}~\bibnamefont{Verresen}},
  \bibinfo{author}{\bibfnamefont{M.~D.} \bibnamefont{Lukin}}, \bibnamefont{and}
  \bibinfo{author}{\bibfnamefont{A.}~\bibnamefont{Vishwanath}},
  \bibinfo{journal}{Phys. Rev. X} \textbf{\bibinfo{volume}{11}},
  \bibinfo{pages}{031005} (\bibinfo{year}{2021}).

\bibitem[{\citenamefont{Yue et~al.}(2021)\citenamefont{Yue, Wang, Mukherjee,
  and Cai}}]{Yue2021}
\bibinfo{author}{\bibfnamefont{M.}~\bibnamefont{Yue}},
  \bibinfo{author}{\bibfnamefont{Z.}~\bibnamefont{Wang}},
  \bibinfo{author}{\bibfnamefont{B.}~\bibnamefont{Mukherjee}},
  \bibnamefont{and} \bibinfo{author}{\bibfnamefont{Z.}~\bibnamefont{Cai}},
  \bibinfo{journal}{Phys. Rev. B} \textbf{\bibinfo{volume}{103}},
  \bibinfo{pages}{L201113} (\bibinfo{year}{2021}).

\bibitem[{\citenamefont{Yan et~al.}(2023)\citenamefont{Yan, Wang, Samajdar,
  Sachdev, and Meng}}]{Yan2023}
\bibinfo{author}{\bibfnamefont{Z.}~\bibnamefont{Yan}},
  \bibinfo{author}{\bibfnamefont{Y.-C.} \bibnamefont{Wang}},
  \bibinfo{author}{\bibfnamefont{R.}~\bibnamefont{Samajdar}},
  \bibinfo{author}{\bibfnamefont{S.}~\bibnamefont{Sachdev}}, \bibnamefont{and}
  \bibinfo{author}{\bibfnamefont{Z.~Y.} \bibnamefont{Meng}},
  \bibinfo{journal}{Phys. Rev. Lett.} \textbf{\bibinfo{volume}{130}},
  \bibinfo{pages}{206501} (\bibinfo{year}{2023}).

\bibitem[{\citenamefont{Sfairopoulos et~al.}(2023)\citenamefont{Sfairopoulos,
  Causer, Mair, and Garrahan}}]{Sfairopoulos2023}
\bibinfo{author}{\bibfnamefont{K.}~\bibnamefont{Sfairopoulos}},
  \bibinfo{author}{\bibfnamefont{L.}~\bibnamefont{Causer}},
  \bibinfo{author}{\bibfnamefont{J.~F.} \bibnamefont{Mair}}, \bibnamefont{and}
  \bibinfo{author}{\bibfnamefont{J.~P.} \bibnamefont{Garrahan}},
  \bibinfo{journal}{Phys. Rev. B} \textbf{\bibinfo{volume}{108}},
  \bibinfo{pages}{174107} (\bibinfo{year}{2023}).

\bibitem[{\citenamefont{{Sfairopoulos}
  et~al.}(2023)\citenamefont{{Sfairopoulos}, {Causer}, {Mair}, and
  {Garrahan}}}]{Sfairopoulos2024}
\bibinfo{author}{\bibfnamefont{K.}~\bibnamefont{{Sfairopoulos}}},
  \bibinfo{author}{\bibfnamefont{L.}~\bibnamefont{{Causer}}},
  \bibinfo{author}{\bibfnamefont{J.~F.} \bibnamefont{{Mair}}},
  \bibnamefont{and} \bibinfo{author}{\bibfnamefont{J.~P.}
  \bibnamefont{{Garrahan}}}, \bibinfo{journal}{arXiv e-prints}
  \bibinfo{eid}{arXiv:2309.08059} (\bibinfo{year}{2023}), \eprint{2309.08059}.

\bibitem[{\citenamefont{Lesanovsky and Garrahan}(2013)}]{Lesanovsky2013}
\bibinfo{author}{\bibfnamefont{I.}~\bibnamefont{Lesanovsky}} \bibnamefont{and}
  \bibinfo{author}{\bibfnamefont{J.~P.} \bibnamefont{Garrahan}},
  \bibinfo{journal}{Phys. Rev. Lett.} \textbf{\bibinfo{volume}{111}},
  \bibinfo{pages}{215305} (\bibinfo{year}{2013}).

\bibitem[{\citenamefont{Lan et~al.}(2018)\citenamefont{Lan, van Horssen,
  Powell, and Garrahan}}]{Lan2018}
\bibinfo{author}{\bibfnamefont{Z.}~\bibnamefont{Lan}},
  \bibinfo{author}{\bibfnamefont{M.}~\bibnamefont{van Horssen}},
  \bibinfo{author}{\bibfnamefont{S.}~\bibnamefont{Powell}}, \bibnamefont{and}
  \bibinfo{author}{\bibfnamefont{J.~P.} \bibnamefont{Garrahan}},
  \bibinfo{journal}{Phys. Rev. Lett.} \textbf{\bibinfo{volume}{121}},
  \bibinfo{pages}{040603} (\bibinfo{year}{2018}).

\bibitem[{\citenamefont{Bluvstein et~al.}(2021)\citenamefont{Bluvstein, Omran,
  Levine, Keesling, Semeghini, Ebadi, Wang, Michailidis, Maskara, Ho
  et~al.}}]{Bluvstein2021}
\bibinfo{author}{\bibfnamefont{D.}~\bibnamefont{Bluvstein}},
  \bibinfo{author}{\bibfnamefont{A.}~\bibnamefont{Omran}},
  \bibinfo{author}{\bibfnamefont{H.}~\bibnamefont{Levine}},
  \bibinfo{author}{\bibfnamefont{A.}~\bibnamefont{Keesling}},
  \bibinfo{author}{\bibfnamefont{G.}~\bibnamefont{Semeghini}},
  \bibinfo{author}{\bibfnamefont{S.}~\bibnamefont{Ebadi}},
  \bibinfo{author}{\bibfnamefont{T.~T.} \bibnamefont{Wang}},
  \bibinfo{author}{\bibfnamefont{A.~A.} \bibnamefont{Michailidis}},
  \bibinfo{author}{\bibfnamefont{N.}~\bibnamefont{Maskara}},
  \bibinfo{author}{\bibfnamefont{W.~W.} \bibnamefont{Ho}},
  \bibnamefont{et~al.}, \bibinfo{journal}{Science}
  \textbf{\bibinfo{volume}{371}}, \bibinfo{pages}{1355} (\bibinfo{year}{2021}).

\bibitem[{\citenamefont{{Wu} et~al.}(2023)\citenamefont{{Wu}, {Wang}, {Yang},
  {Gao}, {Liang}, {Khoon Tey}, {Li}, {Pohl}, and {You}}}]{Wu2023}
\bibinfo{author}{\bibfnamefont{X.}~\bibnamefont{{Wu}}},
  \bibinfo{author}{\bibfnamefont{Z.}~\bibnamefont{{Wang}}},
  \bibinfo{author}{\bibfnamefont{F.}~\bibnamefont{{Yang}}},
  \bibinfo{author}{\bibfnamefont{R.}~\bibnamefont{{Gao}}},
  \bibinfo{author}{\bibfnamefont{C.}~\bibnamefont{{Liang}}},
  \bibinfo{author}{\bibfnamefont{M.}~\bibnamefont{{Khoon Tey}}},
  \bibinfo{author}{\bibfnamefont{X.}~\bibnamefont{{Li}}},
  \bibinfo{author}{\bibfnamefont{T.}~\bibnamefont{{Pohl}}}, \bibnamefont{and}
  \bibinfo{author}{\bibfnamefont{L.}~\bibnamefont{{You}}},
  \bibinfo{journal}{arXiv e-prints} \bibinfo{eid}{arXiv:2305.20070}
  (\bibinfo{year}{2023}), \eprint{2305.20070}.

\bibitem[{\citenamefont{Ates et~al.}(2007)\citenamefont{Ates, Pohl, Pattard,
  and Rost}}]{Ates2007}
\bibinfo{author}{\bibfnamefont{C.}~\bibnamefont{Ates}},
  \bibinfo{author}{\bibfnamefont{T.}~\bibnamefont{Pohl}},
  \bibinfo{author}{\bibfnamefont{T.}~\bibnamefont{Pattard}}, \bibnamefont{and}
  \bibinfo{author}{\bibfnamefont{J.~M.} \bibnamefont{Rost}},
  \bibinfo{journal}{Phys. Rev. Lett.} \textbf{\bibinfo{volume}{98}},
  \bibinfo{pages}{023002} (\bibinfo{year}{2007}).

\bibitem[{\citenamefont{Amthor et~al.}(2010)\citenamefont{Amthor, Giese,
  Hofmann, and Weidem\"uller}}]{Amthor2010}
\bibinfo{author}{\bibfnamefont{T.}~\bibnamefont{Amthor}},
  \bibinfo{author}{\bibfnamefont{C.}~\bibnamefont{Giese}},
  \bibinfo{author}{\bibfnamefont{C.~S.} \bibnamefont{Hofmann}},
  \bibnamefont{and}
  \bibinfo{author}{\bibfnamefont{M.}~\bibnamefont{Weidem\"uller}},
  \bibinfo{journal}{Phys. Rev. Lett.} \textbf{\bibinfo{volume}{104}},
  \bibinfo{pages}{013001} (\bibinfo{year}{2010}).

\bibitem[{\citenamefont{G\"arttner et~al.}(2013)\citenamefont{G\"arttner, Heeg,
  Gasenzer, and Evers}}]{Garttner2013}
\bibinfo{author}{\bibfnamefont{M.}~\bibnamefont{G\"arttner}},
  \bibinfo{author}{\bibfnamefont{K.~P.} \bibnamefont{Heeg}},
  \bibinfo{author}{\bibfnamefont{T.}~\bibnamefont{Gasenzer}}, \bibnamefont{and}
  \bibinfo{author}{\bibfnamefont{J.}~\bibnamefont{Evers}},
  \bibinfo{journal}{Phys. Rev. A} \textbf{\bibinfo{volume}{88}},
  \bibinfo{pages}{043410} (\bibinfo{year}{2013}).

\bibitem[{\citenamefont{Lesanovsky and Garrahan}(2014)}]{Lesanovsky2014}
\bibinfo{author}{\bibfnamefont{I.}~\bibnamefont{Lesanovsky}} \bibnamefont{and}
  \bibinfo{author}{\bibfnamefont{J.~P.} \bibnamefont{Garrahan}},
  \bibinfo{journal}{Phys. Rev. A} \textbf{\bibinfo{volume}{90}},
  \bibinfo{pages}{011603} (\bibinfo{year}{2014}).

\bibitem[{\citenamefont{Marcuzzi et~al.}(2017)\citenamefont{Marcuzzi,
  Min\'a\ifmmode~\check{r}\else \v{r}\fi{}, Barredo, de~L\'es\'eleuc, Labuhn,
  Lahaye, Browaeys, Levi, and Lesanovsky}}]{Marcuzzi2017}
\bibinfo{author}{\bibfnamefont{M.}~\bibnamefont{Marcuzzi}},
  \bibinfo{author}{\bibfnamefont{J.~c.~v.}
  \bibnamefont{Min\'a\ifmmode~\check{r}\else \v{r}\fi{}}},
  \bibinfo{author}{\bibfnamefont{D.}~\bibnamefont{Barredo}},
  \bibinfo{author}{\bibfnamefont{S.}~\bibnamefont{de~L\'es\'eleuc}},
  \bibinfo{author}{\bibfnamefont{H.}~\bibnamefont{Labuhn}},
  \bibinfo{author}{\bibfnamefont{T.}~\bibnamefont{Lahaye}},
  \bibinfo{author}{\bibfnamefont{A.}~\bibnamefont{Browaeys}},
  \bibinfo{author}{\bibfnamefont{E.}~\bibnamefont{Levi}}, \bibnamefont{and}
  \bibinfo{author}{\bibfnamefont{I.}~\bibnamefont{Lesanovsky}},
  \bibinfo{journal}{Phys. Rev. Lett.} \textbf{\bibinfo{volume}{118}},
  \bibinfo{pages}{063606} (\bibinfo{year}{2017}).

\bibitem[{\citenamefont{P\'erez-Espigares
  et~al.}(2017)\citenamefont{P\'erez-Espigares, Marcuzzi, Guti\'errez, and
  Lesanovsky}}]{Espigares2017}
\bibinfo{author}{\bibfnamefont{C.}~\bibnamefont{P\'erez-Espigares}},
  \bibinfo{author}{\bibfnamefont{M.}~\bibnamefont{Marcuzzi}},
  \bibinfo{author}{\bibfnamefont{R.}~\bibnamefont{Guti\'errez}},
  \bibnamefont{and}
  \bibinfo{author}{\bibfnamefont{I.}~\bibnamefont{Lesanovsky}},
  \bibinfo{journal}{Phys. Rev. Lett.} \textbf{\bibinfo{volume}{119}},
  \bibinfo{pages}{140401} (\bibinfo{year}{2017}).

\bibitem[{\citenamefont{Causer et~al.}(2020)\citenamefont{Causer, Lesanovsky,
  Ba\~nuls, and Garrahan}}]{Causer2020}
\bibinfo{author}{\bibfnamefont{L.}~\bibnamefont{Causer}},
  \bibinfo{author}{\bibfnamefont{I.}~\bibnamefont{Lesanovsky}},
  \bibinfo{author}{\bibfnamefont{M.~C.} \bibnamefont{Ba\~nuls}},
  \bibnamefont{and} \bibinfo{author}{\bibfnamefont{J.~P.}
  \bibnamefont{Garrahan}}, \bibinfo{journal}{Phys. Rev. E}
  \textbf{\bibinfo{volume}{102}}, \bibinfo{pages}{052132}
  (\bibinfo{year}{2020}).

\bibitem[{\citenamefont{Helmrich et~al.}(2020)\citenamefont{Helmrich, Arias,
  Lochead, Wintermantel, Buchhold, Diehl, and Whitlock}}]{Helmrich2020}
\bibinfo{author}{\bibfnamefont{S.}~\bibnamefont{Helmrich}},
  \bibinfo{author}{\bibfnamefont{A.}~\bibnamefont{Arias}},
  \bibinfo{author}{\bibfnamefont{G.}~\bibnamefont{Lochead}},
  \bibinfo{author}{\bibfnamefont{T.~M.} \bibnamefont{Wintermantel}},
  \bibinfo{author}{\bibfnamefont{M.}~\bibnamefont{Buchhold}},
  \bibinfo{author}{\bibfnamefont{S.}~\bibnamefont{Diehl}}, \bibnamefont{and}
  \bibinfo{author}{\bibfnamefont{S.}~\bibnamefont{Whitlock}},
  \bibinfo{journal}{Nature} \textbf{\bibinfo{volume}{577}},
  \bibinfo{pages}{481} (\bibinfo{year}{2020}).

\bibitem[{\citenamefont{Ding et~al.}(2020)\citenamefont{Ding, Busche, Shi, Guo,
  and Adams}}]{Ding2020}
\bibinfo{author}{\bibfnamefont{D.-S.} \bibnamefont{Ding}},
  \bibinfo{author}{\bibfnamefont{H.}~\bibnamefont{Busche}},
  \bibinfo{author}{\bibfnamefont{B.-S.} \bibnamefont{Shi}},
  \bibinfo{author}{\bibfnamefont{G.-C.} \bibnamefont{Guo}}, \bibnamefont{and}
  \bibinfo{author}{\bibfnamefont{C.~S.} \bibnamefont{Adams}},
  \bibinfo{journal}{Phys. Rev. X} \textbf{\bibinfo{volume}{10}},
  \bibinfo{pages}{021023} (\bibinfo{year}{2020}).

\bibitem[{\citenamefont{Liu et~al.}(2022)\citenamefont{Liu, Yang, Bienias,
  Iadecola, and Gorshkov}}]{Liu2022}
\bibinfo{author}{\bibfnamefont{F.}~\bibnamefont{Liu}},
  \bibinfo{author}{\bibfnamefont{Z.-C.} \bibnamefont{Yang}},
  \bibinfo{author}{\bibfnamefont{P.}~\bibnamefont{Bienias}},
  \bibinfo{author}{\bibfnamefont{T.}~\bibnamefont{Iadecola}}, \bibnamefont{and}
  \bibinfo{author}{\bibfnamefont{A.~V.} \bibnamefont{Gorshkov}},
  \bibinfo{journal}{Phys. Rev. Lett.} \textbf{\bibinfo{volume}{128}},
  \bibinfo{pages}{013603} (\bibinfo{year}{2022}).

\bibitem[{\citenamefont{Sandvik}(2003)}]{Sandvik2003}
\bibinfo{author}{\bibfnamefont{A.~W.} \bibnamefont{Sandvik}},
  \bibinfo{journal}{Phys. Rev. E} \textbf{\bibinfo{volume}{68}},
  \bibinfo{pages}{056701} (\bibinfo{year}{2003}).

\bibitem[{\citenamefont{{Merali} et~al.}(2021)\citenamefont{{Merali}, {De
  Vlugt}, and {Melko}}}]{Merali2021}
\bibinfo{author}{\bibfnamefont{E.}~\bibnamefont{{Merali}}},
  \bibinfo{author}{\bibfnamefont{I.~J.~S.} \bibnamefont{{De Vlugt}}},
  \bibnamefont{and} \bibinfo{author}{\bibfnamefont{R.~G.}
  \bibnamefont{{Melko}}}, \bibinfo{journal}{arXiv e-prints}
  \bibinfo{eid}{arXiv:2107.00766} (\bibinfo{year}{2021}), \eprint{2107.00766}.

\bibitem[{\citenamefont{{Patil}}(2023)}]{Patil2023}
\bibinfo{author}{\bibfnamefont{P.}~\bibnamefont{{Patil}}},
  \bibinfo{journal}{arXiv e-prints} \bibinfo{eid}{arXiv:2309.00482}
  (\bibinfo{year}{2023}), \eprint{2309.00482}.

\bibitem[{\citenamefont{Rau and Gingras}(2016)}]{Rau2016}
\bibinfo{author}{\bibfnamefont{J.~G.} \bibnamefont{Rau}} \bibnamefont{and}
  \bibinfo{author}{\bibfnamefont{M.~J.~P.} \bibnamefont{Gingras}},
  \bibinfo{journal}{Nat. Comms} \textbf{\bibinfo{volume}{7}},
  \bibinfo{pages}{12234} (\bibinfo{year}{2016}).

\bibitem[{\citenamefont{Schempp et~al.}(2014)\citenamefont{Schempp, G\"unter,
  Robert-de Saint-Vincent, Hofmann, Breyel, Komnik, Sch\"onleber, G\"arttner,
  Evers, Whitlock et~al.}}]{Schempp2014}
\bibinfo{author}{\bibfnamefont{H.}~\bibnamefont{Schempp}},
  \bibinfo{author}{\bibfnamefont{G.}~\bibnamefont{G\"unter}},
  \bibinfo{author}{\bibfnamefont{M.}~\bibnamefont{Robert-de Saint-Vincent}},
  \bibinfo{author}{\bibfnamefont{C.~S.} \bibnamefont{Hofmann}},
  \bibinfo{author}{\bibfnamefont{D.}~\bibnamefont{Breyel}},
  \bibinfo{author}{\bibfnamefont{A.}~\bibnamefont{Komnik}},
  \bibinfo{author}{\bibfnamefont{D.~W.} \bibnamefont{Sch\"onleber}},
  \bibinfo{author}{\bibfnamefont{M.}~\bibnamefont{G\"arttner}},
  \bibinfo{author}{\bibfnamefont{J.}~\bibnamefont{Evers}},
  \bibinfo{author}{\bibfnamefont{S.}~\bibnamefont{Whitlock}},
  \bibnamefont{et~al.}, \bibinfo{journal}{Phys. Rev. Lett.}
  \textbf{\bibinfo{volume}{112}}, \bibinfo{pages}{013002}
  (\bibinfo{year}{2014}).

\bibitem[{\citenamefont{Urvoy et~al.}(2015)\citenamefont{Urvoy, Ripka,
  Lesanovsky, Booth, Shaffer, Pfau, and L\"ow}}]{Urvoy2015}
\bibinfo{author}{\bibfnamefont{A.}~\bibnamefont{Urvoy}},
  \bibinfo{author}{\bibfnamefont{F.}~\bibnamefont{Ripka}},
  \bibinfo{author}{\bibfnamefont{I.}~\bibnamefont{Lesanovsky}},
  \bibinfo{author}{\bibfnamefont{D.}~\bibnamefont{Booth}},
  \bibinfo{author}{\bibfnamefont{J.~P.} \bibnamefont{Shaffer}},
  \bibinfo{author}{\bibfnamefont{T.}~\bibnamefont{Pfau}}, \bibnamefont{and}
  \bibinfo{author}{\bibfnamefont{R.}~\bibnamefont{L\"ow}},
  \bibinfo{journal}{Phys. Rev. Lett.} \textbf{\bibinfo{volume}{114}},
  \bibinfo{pages}{203002} (\bibinfo{year}{2015}).

\bibitem[{\citenamefont{Zadnik and Garrahan}(2023)}]{Zadnik2023}
\bibinfo{author}{\bibfnamefont{L.}~\bibnamefont{Zadnik}} \bibnamefont{and}
  \bibinfo{author}{\bibfnamefont{J.~P.} \bibnamefont{Garrahan}},
  \bibinfo{journal}{Phys. Rev. B} \textbf{\bibinfo{volume}{108}},
  \bibinfo{pages}{L100304} (\bibinfo{year}{2023}).

\bibitem[{Sup()}]{Supplementary}
\bibinfo{howpublished}{See the supplementary material for an elaboration of the
  fractal cluster dimension, as well as a comparision between the ground states
  within the single-cluster subspace and subspaces with other cluster numbers.}

\bibitem[{\citenamefont{Vicsek}(1992)}]{Vicsek1992}
\bibinfo{author}{\bibfnamefont{T.}~\bibnamefont{Vicsek}},
  \emph{\bibinfo{title}{Fractal Growth Phenomena, Part IV}}
  (\bibinfo{publisher}{World Scientific, Singapore}, \bibinfo{year}{1992}).

\bibitem[{\citenamefont{Hallen et~al.}(2022)\citenamefont{Hallen, Grigera,
  Tennant, Castelnovo, and Moessner}}]{Hallen2022}
\bibinfo{author}{\bibfnamefont{J.~N.} \bibnamefont{Hallen}},
  \bibinfo{author}{\bibfnamefont{S.~A.} \bibnamefont{Grigera}},
  \bibinfo{author}{\bibfnamefont{D.~A.} \bibnamefont{Tennant}},
  \bibinfo{author}{\bibfnamefont{C.}~\bibnamefont{Castelnovo}},
  \bibnamefont{and} \bibinfo{author}{\bibfnamefont{R.}~\bibnamefont{Moessner}},
  \bibinfo{journal}{Science} \textbf{\bibinfo{volume}{378}},
  \bibinfo{pages}{1218} (\bibinfo{year}{2022}).

\bibitem[{\citenamefont{Feder}(1988)}]{Feder1988}
\bibinfo{author}{\bibfnamefont{J.}~\bibnamefont{Feder}},
  \emph{\bibinfo{title}{Fractals}} (\bibinfo{publisher}{Springer Science, New
  York}, \bibinfo{year}{1988}).

\bibitem[{\citenamefont{Garrahan}(2018)}]{Garrahan2018}
\bibinfo{author}{\bibfnamefont{J.~P.} \bibnamefont{Garrahan}},
  \bibinfo{journal}{Physica A: Statistical Mechanics and its Applications}
  \textbf{\bibinfo{volume}{504}}, \bibinfo{pages}{130} (\bibinfo{year}{2018}).

\bibitem[{\citenamefont{Mezard et~al.}(1986)\citenamefont{Mezard, Parisi, and
  Virasoro}}]{Mezard1986}
\bibinfo{author}{\bibfnamefont{M.}~\bibnamefont{Mezard}},
  \bibinfo{author}{\bibfnamefont{G.}~\bibnamefont{Parisi}}, \bibnamefont{and}
  \bibinfo{author}{\bibfnamefont{M.}~\bibnamefont{Virasoro}},
  \emph{\bibinfo{title}{Spin Glass Theory and Beyond: An Introduction to the
  Replica Method and Its Applications}} (\bibinfo{publisher}{~World
  Scientific}, \bibinfo{year}{1986}).

\bibitem[{\citenamefont{Giamarchi and Schulz}(1988)}]{Giamarchi1988}
\bibinfo{author}{\bibfnamefont{T.}~\bibnamefont{Giamarchi}} \bibnamefont{and}
  \bibinfo{author}{\bibfnamefont{H.~J.} \bibnamefont{Schulz}},
  \bibinfo{journal}{Phys. Rev. B} \textbf{\bibinfo{volume}{37}},
  \bibinfo{pages}{325} (\bibinfo{year}{1988}).

\bibitem[{\citenamefont{Fisher et~al.}(1989)\citenamefont{Fisher, Weichman,
  Grinstein, and Fisher}}]{Fisher1989}
\bibinfo{author}{\bibfnamefont{M.~P.~A.} \bibnamefont{Fisher}},
  \bibinfo{author}{\bibfnamefont{P.~B.} \bibnamefont{Weichman}},
  \bibinfo{author}{\bibfnamefont{G.}~\bibnamefont{Grinstein}},
  \bibnamefont{and} \bibinfo{author}{\bibfnamefont{D.~S.}
  \bibnamefont{Fisher}}, \bibinfo{journal}{Phys. Rev. B}
  \textbf{\bibinfo{volume}{40}}, \bibinfo{pages}{546} (\bibinfo{year}{1989}).

\bibitem[{\citenamefont{Hermele et~al.}(2004)\citenamefont{Hermele, Senthil,
  Fisher, Lee, Nagaosa, and Wen}}]{Hermele2004}
\bibinfo{author}{\bibfnamefont{M.}~\bibnamefont{Hermele}},
  \bibinfo{author}{\bibfnamefont{T.}~\bibnamefont{Senthil}},
  \bibinfo{author}{\bibfnamefont{M.~P.~A.} \bibnamefont{Fisher}},
  \bibinfo{author}{\bibfnamefont{P.~A.} \bibnamefont{Lee}},
  \bibinfo{author}{\bibfnamefont{N.}~\bibnamefont{Nagaosa}}, \bibnamefont{and}
  \bibinfo{author}{\bibfnamefont{X.-G.} \bibnamefont{Wen}},
  \bibinfo{journal}{Phys. Rev. B} \textbf{\bibinfo{volume}{70}},
  \bibinfo{pages}{214437} (\bibinfo{year}{2004}).

\bibitem[{\citenamefont{Henley}(2010)}]{Henley2010}
\bibinfo{author}{\bibfnamefont{C.}~\bibnamefont{Henley}},
  \bibinfo{journal}{Annu. Rev. Condens. Mat. Phys.}
  \textbf{\bibinfo{volume}{1}}, \bibinfo{pages}{179} (\bibinfo{year}{2010}).

\bibitem[{\citenamefont{C.Castelnovo et~al.}(2012)\citenamefont{C.Castelnovo,
  R.Moessner, and Sondhi}}]{Castelnovo2012}
\bibinfo{author}{\bibnamefont{C.Castelnovo}},
  \bibinfo{author}{\bibnamefont{R.Moessner}}, \bibnamefont{and}
  \bibinfo{author}{\bibfnamefont{S.}~\bibnamefont{Sondhi}},
  \bibinfo{journal}{Annu. Rev. Condens. Mat. Phys.}
  \textbf{\bibinfo{volume}{3}}, \bibinfo{pages}{35} (\bibinfo{year}{2012}).

\bibitem[{\citenamefont{Kosterlitz and Thouless}(1973)}]{Kosterlitz1973}
\bibinfo{author}{\bibfnamefont{J.~M.} \bibnamefont{Kosterlitz}}
  \bibnamefont{and} \bibinfo{author}{\bibfnamefont{D.~J.}
  \bibnamefont{Thouless}}, \bibinfo{journal}{Journal of Physics C: Solid State
  Physics} \textbf{\bibinfo{volume}{6}}, \bibinfo{pages}{1181}
  (\bibinfo{year}{1973}).

\end{thebibliography}

\end{document}